\documentclass[12pt]{article}
 \usepackage{epsf}
 \usepackage{epsfig}
 \def\beq{\begin{eqnarray}}
 \def\eeq{\end{eqnarray}}
 \def\beqs{\begin{eqnarray*}}
 \def\eeqs{\end{eqnarray*}}
 \def\dl{\delta}
 \newcommand{\be}{\begin{equation}}
 \newcommand{\ee}{\end{equation}}
 \newcommand{\lll}{\langle}
 \newcommand{\rrr}{\rangle}
 
 \newcommand{\llll}{\langle\langle}
 \newcommand{\rrrr}{\rangle\rangle}
 \newcommand{\T}{\mbox{Tr}\> }

 \def\centeron#1#2{{\setbox0=\hbox{#1}\setbox1=\hbox{#2}\ifdim
 \wd1>\wd0\kern.5\wd1\kern-.5\wd0\fi
 \copy0\kern-.5\wd0\kern-.5\wd1\copy1\ifdim\wd0>\wd1
 \kern.5\wd0\kern-.5\wd1\fi}}
 \def\ltap{\;\centeron{\raise.35ex\hbox{$<$}}{\lower.65ex\hbox{$\sim$}}\;}
 \def\gtap{\;\centeron{\raise.35ex\hbox{$>$}}{\lower.65ex\hbox{$\sim$}}\;}
 \def\gsim{\mathrel{\gtap}}

\begin{document}
 \begin{titlepage}
 \begin{flushright}
 {ITP-UU-02/17}
 \end{flushright}

 \vskip 1.2cm

 \begin{center}

 {\LARGE\bf Interaction of Wilson loops \\
 in confining vacuum
 }
 \vskip 1.4cm
 {\bf \large  V.I. Shevchenko}$^{1,3}$, {\bf\large Yu.A.Simonov}$^{2,3}$.
 \\
 \vskip 0.3cm
 $^1${\it Institute for Theoretical Physics, Utrecht University,
 \\ Utrecht, the Netherlands}
 \\
 \vskip 0.2cm
 $^2${\it Jefferson Laboratory \\  Newport News, VA, USA }
 \\
 \vskip 0.2cm
 $^3${\it Institute of Theoretical and Experimental Physics,
 \\ Moscow, Russia } \\
 \vskip 0.25cm
 e-mail: V.Shevchenko@phys.uu.nl; simonov@heron.itep.ru

 \vskip 2cm

 \begin{abstract}

Nonperturbative and perturbative interaction mechanisms of Wilson loops
 are studied within the background field formalism. The first one
operates when distance between minimal surfaces of the loops is
small and may be important for sea quark effects and strong decay
processes. The second mechanism -- perturbative interaction in
nonperturbative confining background is found to be physically
dominant for all loop configurations characteristic of scattering
process. It reduces to perturbative gluon exchanges at small
distances, while at larger distances it corresponds to the
$t$-channel exchange of glueball states.
 Comparison to other approaches is made and possible physical
 applications are discussed.

 \end{abstract}
 \end{center}

 \vskip 1.0 cm

 \end{titlepage}

 \setcounter{footnote}{0} \setcounter{page}{2}
 \setcounter{section}{0} \setcounter{subsection}{0}
 \setcounter{subsubsection}{0}


 \section{Introduction}

 Interaction of Wilson loops in QCD is the basic element of many
 physical applications. One can mention hadron-hadron
 scattering amplitude, in particular the phenomenology of Pomeron exchange;
 the assumed color transparency phenomenon; strong hadron decays
 and OZI-forbidden processes etc. There is also considerable interest
in calculating Wilson loop and Polyakov loop correlators {\it per
se}, not only in QCD but also in other field theories, in
particular in supersymmetric Yang-Mills theory.
 In all cases one starts with connected average of two (or more)
Wilson
 loops and tries to calculate it in a kinematic region of interest using
appropriate field-theoretical technique.
   It is the aim of the
present
 paper to do it in the framework of field correlator method in
gluodynamics (see, e.g.
 review \cite{ddss} and references therein) incorporating both
perturbative
 and nonperturbative contributions.
 Let us briefly remind the basic ideas behind the method. The general
Wilson
 loop approach was introduced originally for heavy quarkonia \cite{w}.
 For quarks of finite mass one can use Feynman-Schwinger representation
 (see \cite{ffs} for a review and references therein) to
 write the meson Green's function as an integral over all possible Wilson
 loops, formed by the quark trajectories and finally to express the meson
 (and baryon) dynamics in terms of gauge-invariant correlators of the field
 strengths,
 characterizing the properties of confining background. When going to
 the hadron-hadron scattering one can adopt the same formalism to express
 scattering amplitude through the vacuum average of the product of two
Wilson
 loops, with subsequent integration over all ensemble of loops.
 In doing so we use the background field formalism to separate
 nonperturbative
 gluon configurations from perturbative (sometimes called "valence")
gluons.
 Thus the answer will contain two parts: purely nonperturbative and
 perturbative inside noperturbative background, i.e. glueball exchanges
 between Wilson loops.

 For the former it is convenient to use nonabelian
 Stokes theorem and express the answer in terms of the gauge-invariant
field correlators and finally via the string tension. Perturbative part in the
 nonperturbative background corresponds to exchange of glueball states between loops.
 We shall keep number of colors $N_c$ as a free parameter in what follows.
 It will be argued that the leading term for a typical kinematics of the
 scattering process is the (background-modified) perturbative one.
 This is in line with the old observation that high-energy scattering
 amplitude is dominated by the Pomeron exchange.
 We do not consider here the leading in $1/N_c$ terms of ordinary Reggeon
 exchanges, which formally refer to one-loop case and are subleading in the
 high energy limit. We are also not discussing pion exchanges
 which may give the main contribution in some cases at not large energies,
  and will concentrate our attention
 on the case of theory without dynamical quarks, i.e. gluodynamics.
In another physical situation, e.g. when accounting for the
sea-quark loops or
 for a decay transition of a hadron state, the roles of perturbative and
 nonperturbative mechanisms may change depending on the hadron quantum numbers.
 In all cases however the loop-loop interaction is the starting point of
 field-correlator formalism application for scattering, strong decay etc.
  In section 2 we introduce the general background formalism for the
  interaction of Wilson loops. In section 3 the nonperturbative mechanism
  is studied in detail. In section 4 perturbative gluon exchange is shown
  to transform into glueball exchange mechanism at large distances.
  Section 5 is devoted to a physical discussion of results and brief comparison
  with the existing models.

 \section
 {Interaction of Wilson loops in the background field formalism}

 In this section we are going to exploit the background field formalism
 \cite{back} in the form worked out in \cite{si1}.
 We refer the interested
 reader to the cited papers for all the details and recall basic
 steps only briefly.
 We start with decomposing of gluon field $A_{\mu}(x)$
 into nonperturbative background $B_{\mu}(x)$ and perturbative part
 $a_{\mu}$, propagating in the background:
 \be
 A_{\mu} = B_{\mu} + a_{\mu}
 \label{e1}
 \ee
 Total gauge transformation is decomposed as
 \be
 B_{\mu} \to U^{\dagger} \left( B_{\mu} - \frac{i}{g} \partial_{\mu}
\right)
 U \;\; ; \;\; a_{\mu} \to U^{\dagger} a_{\mu} U
 \label{e2}
 \ee
 The principle of separation is of no importance at the level of
 partition function due to obvious identity
 \be
 Z= \frac{1}{{\cal N}} \int {\cal D} A_{\mu} \exp (-S[A]) =
  \frac{1}{{\cal N}'} \int {\cal D} B_{\mu} \int {\cal D} a_{\mu} \exp
 (-S[B+a])
 \label{e3}
 \ee
 (here gauge-fixing and ghost terms are assumed to be included into the
 measure of integration).
 The Wilson loop depends on both $B_{\mu}$ and $a_{\mu}$ :
 $$
 W(C) = \frac{1}{N_c}\>\T {\mbox P}\exp\left(ig\int\limits_{C}
 (B_{\mu}^c + a_{\mu}^c) t^c dz_{\mu}\right) =
 $$
 \be
 = \frac{1}{N_c} \lim\limits_{M\to\infty} \T \prod\limits_{m=1}^{M} {\mbox
P}
 \left(1+ig(B_{\mu}(z^{[m]}) + a_{\mu}(z^{[m]}))\Delta z_{\mu}^{[m]}\right)
 \label{e5}
 \ee
 The trace in fundamental representation is normalized as
 $$
 \T {\hat{\bf 1}} = N_c \;\;\; ; \;\;\;
 \T t^a t^b = \frac12 {\delta}^{ab}
 $$

 Our general strategy is the following \cite{si1}:
 we expand the correlators under study in powers of the field
 $a_{\mu}$, while account for effects caused by the
 nonperturbative background exactly (i.e. without expansion
 in powers of $B_{\mu}$). Namely, one has
 \be
 W(C) = W^{(0)}(C) + W^{(1)}(C) + W^{(2)}(C) + ...
 \label{e6}
 \ee
 where, e.g.
 \be
 W^{(1)}(C) = \frac{ig}{N_c}\> \T {\mbox P}_{zu} \int\limits_{C} a_{\mu}(z)
 dz_{\mu} \>
 \exp\left(ig\int\limits_{C_{z}}
 B_{\nu}(u) du_{\nu}\right)
 \label{e7}
 \ee
 while $W^{(0)}(C)$ contains only the field $B_{\mu}$ and
 ordering operator ${\mbox P}_{zu} $ takes care of
 ordering $a_{\mu}(z)$ and $B_{\mu}(u)$.

 Let us define now connected average of two Wilson loops
 as
 \be
 \chi(C_1, C_2) = \lll W(C_1) W(C_2) \rrr -
 \lll W(C_1)\rrr\lll W(C_2) \rrr
 \label{e8}
 \ee
 This average can also be expanded in powers of $ga_{\mu}$:

 \be
 \chi(C_1, C_2) =
 \chi^{(0)}(C_1, C_2) + \chi^{(2)}(C_1, C_2) +  \chi^{(4)}(C_1, C_2) +...
 \label{e9}
 \ee
 Here $\chi^{(0)}(C_1, C_2) $ is purely nonperturbative interaction
 of two Wilson loops and depends only on fields $B_{\mu}$, while
 higher terms $\chi^{(n)}(C_1, C_2)$  are proportional to the average of
 $(ga_{\mu})^n$. One immediately notices that since $\lll W^{(1)} \rrr$
 is identically zero, the term $\chi^{(2)}(C_1, C_2) $ vanishes
 and the expansion starts with the two-gluon exchange term $\chi^{(4)}(C_1,
 C_2)$.

 In some cases also $C$-odd exchange contribution (odderon-type) is
 important,
 it is contained in $\chi^{(6)}(C_1, C_2)$. In what follows we discuss
 mostly the purely nonperturbative term $\chi^{(0)}(C_1, C_2) $,
 and in the last part of the paper also two-gluon exchange
 $\chi^{(4)}(C_1, C_2)$ in section 4.

 \section{Nonperturbative interaction of Wilson loops}

 We consider in this section the first term in the expansion
 (\ref{e9}), namely $\chi^{(0)}(C_1, C_2) $ and use the contour gauge
 \cite{cont} to write down the Wilson loop as surface integral
 \be
 W(C) = \frac{1}{N_c}\>\T {\mbox P}\exp\left(ig\int\limits_{S}
 F_{\mu\nu}(u,x_0) d\sigma_{\mu\nu}(u)\right)
 \label{e10}
 \ee
 We have defined in (\ref{e10})
 $
 F_{\mu\nu}(u,x_0) = \Phi_{L_{x_0 u}}F_{\mu\nu}(u)\Phi_{L_{u x_0}}
 $
 where phase factors along the curve $L_{x_0 u}$ with the edge points $x_0$ and $u$
 are given by
 \be
 \Phi_{L_{x_0 u}}= {\mbox P}\exp\left(ig\int\limits_{u}^{x_0}
 B_{\mu}(z) dz{\mu}\right)
 \label{e12}
 \ee

 Consider Wilson loops defined for two contours $C_1$ and $C_2$,
 where individual minimal surfaces will be denoted as
 $S_1^{min}$ and $S_2^{min}$,
respectively throughout the paper.\footnote{For any surface we use
one and the same letter $S$ for
 a surface as geometrical object and for its area.}
The typical problem in the discussed framework is to choose
optimal integration surfaces in the integrals of the form
(\ref{e10}). The Wilson loop (\ref{e10}) is gauge-invariant and
 surface-independent. In the confining regime it obeys the area law,
 which means that a single surface is unambiguously chosen dynamically
 (modulo surface fluctuations) and it is naturally assumed to be the
 minimal surface. In the approach proposed in \cite{ds} one can visualize
 how the minimal surface results from the balance of action of all
 correlators, and higher correlators play important role in making the
 surface smooth and minimal. After this goal is achieved, the relative
 importance of correlators on the minimal surface obeys a strict hierarchy.
For a single Wilson loop it can be strongly argued (see discussion
and references in \cite{ddss}, \cite{casimir}),
 that in case of minimal surface (which is obviously the distinguished
surface for given contour) the dominant nonperturbative
contribution in the cluster expansion of (\ref{e10}) comes from
the lowest, Gaussian correlator of the field strength operators.
This property is known as Gaussian dominance \cite{ds} and it
plays important role in all phenomenological applications of field
correlator method.

In the physical picture described above appearance of the minimal
surface has been in some sense a result of field correlator
dynamics. However, one can take another view by saying that
Gaussian dominance corresponds to such profile of the confining
string worldsheet, which minimizes the total energy of the system.
In other words, to calculate $\lll W(C) \rrr$ one may proceed as
follows: first, to find the minimal energy of confining string
configuration (trivially corresponding to minimal area surface in
case of single static loop) and, as the next step, to calculate
average (\ref{e10}) with Gaussian ensemble of correlators
integrated over this surface. It is this principle which we shall
use in what follows to choose the shape of the surfaces entering
our problem. To illustrate the physics behind it, let us consider
gauge-invariant Green's function of scalar particle of the mass
$M_1$ and antiparticle of the mass $M_2$ interacting with
nonabelian gauge field $A_{\mu}^a$ (see, e.g. \cite{ffs}) $$
 G^{[2]}(x,y) = \lll \phi^{\dagger}(x) \Phi_{L_{x\bar x}} \phi(\bar x) \phi^{\dagger}(\bar y)
 \Phi_{L_{\bar y y}} \phi(y) \rrr =
 $$
 \be
 =
 \int\limits_0^{\infty} ds_1 \int\limits_0^{\infty} ds_2
 \int \left({\cal D} z \right)_{xy} \int \left({\cal D} {\bar z} \right)_{\bar x\bar y}
 \>\exp (-K_0) \lll W (C_{xy\bar y \bar x}) \rrr
 \label{e45}
 \ee
 where
 \be
 K_{0} = M_1^2 s_1 + M_2^2 s_2 +
 \frac14 \int\limits_0^{s_1} d\tau_1 \left(
 \frac{d z_{\mu}}{d\tau_1}\right)^2 +
 \frac14 \int\limits_0^{s_2} d\tau_2 \left(
 \frac{d {\bar z}_{\mu}}{d\tau_2}\right)^2
 \label{y7}
 \ee
and the contour $C_{xy\bar y \bar x}$ is formed by dynamical
trajectories $z, {\bar z}$ and the lines $ L_{x \bar x}, L_{\bar y
y}$. Green's function (\ref{e45}) encodes all information about
the spectrum of the system. Of special interest are particular
cases when some trajectories in the integrals in (\ref{e45}) are
singled out kinematically as providing the main contribution.
Phenomenologically the most important cases are represented by the
large mass (nonrelativistic) limit, where the particle
trajectories are close to parallel to the temporal (Euclidean)
axis, and large-momentum limit, where eikonal approximation is
applicable. Let us consider the former case in more details and
assume for simplicity that $x_4 = {\bar x}_4$, $y_4 = {\bar y}_4$,
$y_4 - x_4 =T$, and center-of-mass motion of the considered
two--body system is absent. The Green's function (\ref{e45}) can
be spectrally decomposed as \be G^{[2]}(x , \bar x ; y, \bar y ) =
\sum\limits_{\{n\}} \psi^{\dagger}_{\{ n \}} (\vec{x} - \vec{\bar
x}) \psi_{\{ n \} } (\vec{y} - \vec{\bar y}) \exp(-E^{\{n\}} T )
\label{sur} \ee where $T$ stays for the Euclidean time interval
while $\psi_{\{ n \} } , E^{\{ n \} }$ denote wave function and
energy of the state with quantum numbers ${\{ n \} }$,
respectively. Comparing (\ref{e45}) and (\ref{sur}) we notice that
the problem of minimizing energy levels is in close correspondence
with the problem of maximizing average value of the Wilson loop
$\lll  W(C) \rrr$ for given contour, prescribed by kinematics.
This correspondence becomes straightforward in classical limit of
static rectangular loop where area law dictates
$$
\lll W(R\times T) \rrr \approx \exp(-ET)
$$
and the static energy is given by $E=\sigma R$, which corresponds
to straight line string and minimal area. Notice that we confine
ourselves to the leading area law terms in this paper and do not
consider perimeter contributions.

 For two or more Wilson loops the situation is more
complicated, but general framework is the same. Indeed, one can
consider multi-point Green's functions, e.g. two--meson one \be
G^{[4]} = \left\lll \Psi^{\dagger}_{L_{x {\bar x}}}
\Psi^{\dagger}_{L_{u {\bar u}}} \Psi_{L_{y {\bar y}}}\Psi_{L_{v
{\bar v}}} \right\rrr \label{g4} \ee
 where
$$
\Psi_{L_{y {\bar y}}} =   \phi^{\dagger}(y) \Phi_{L_{y \bar y}}
\phi(\bar y)
$$
and again adopt Feynman-Schwinger representation to get \be
G^{[4]} \sim \int [***] \cdot \lll W(C_1) W(C_2) \rrr + ...
\label{g4w} \ee where stars $[***]$ stay for integration measure
while dots denote other possible ways to contract field operators
in the definition (\ref{g4}) and get, correspondingly, other
geometries of contours $C_1$, $C_2$. On the other hand, the
Green's function (\ref{g4}) describing interacting two--meson
system, must be dominated at large separation/time by the
confining strings configuration of lowest total energy, in direct
analogy with one meson case.

It is physically obvious that for two well separated loops the
minimal configuration of confining strings is given by two
individual minimal surfaces with no common points of intersection.
From the point of view of the expression (\ref{g4w}) it means that
average $\lll W(C_1) W(C_2)\rrr$ factorizes into $\lll
W(C_1)\rrr\lll W(C_2)\rrr$ and no nonperturbative interaction
takes place. Notice that perturbative interaction is mediated in
our picture by propagating degrees of freedom, while $\chi^{(0)}$
could receive nonzero contributions only because of rearrangement
of confining strings with respect to the free case, if it happens.
In the approach developed in the present paper it is always
assumed that the gluon correlation length $T_g$ (which also
defines the radius of the string) is taken vanishingly small,
while the string tension $\sigma$ is kept fixed and therefore
$\chi^{(0)}$ is expressed in terms of $\sigma$ alone. This is in
contrast to another approach \cite{doschn}, to be discussed later
in the paper.

Consider now a process of decreasing the distance between the
loops. For classical films, as it is easily checked in experiments
with soap films, at some point there appears the situation, when
the minimal surface, bounded by contours $C_1$ and $C_2$, which we
shall call $S_{12}$, becomes smaller, than the sum of $S^{min}_1$
and $S^{min}_2$. At this point, according to minimum energy
principle, there should appear a common surface $S_{12}$ which
constitutes the interaction between the loops. Indeed the
classical films (e.g. soap films) satisfy the stated above
principle. For the simplest case of two concentric circles of
equal radius $R$ at distance $h$ it was Leonard Euler who
demonstrated that the corresponding minimal surface $S_{12}$ is
a catenoid and it exists for any $h$ such that the equation \be
\frac{R}{R_{min}} = \cosh \left( \frac{h}{2R_{min}} \right)
\label{er5} \ee has a solution. Here $R_{min}$ defines the so
called minimal radius of the catenoid. At some critical distance
$h_{cr}$ the solution ceases to exist and the minimal surface
coincides with two disconnected disks.\footnote{Actually the
analog of condition (\ref{en}) is violated earlier, at some
distance ${\tilde h}_{cr} < h_{cr}$. The catenoid solution is
locally stable but globally unstable for ${\tilde h}_{cr} < h <
h_{cr}$. At $h > h_{cr}$ local stability is lost. We do not make a
distinction between ${\tilde h}_{cr}$ and $h_{cr}$ in our
consideration and speaking about critical distance we always have
in mind minimal energy condition.}

The case of QCD is more complicated, first of all because of
nonabelian nature of the gluon field. In particular, it is possible
strictly speaking to take in (\ref{e10}) integration
surface with the topology different from that of the disc only at
the price of introducing additional holonomy
factors.\footnote{See, e.g. \cite{jap}. It is in contrast with
abelian Stokes theorem where one can easily integrate two-form
field strength over noncontractible surface.} It is also worth
mentioning that
 the surface corresponding to the
global minimum of energy can be glued with itself in nontrivial
way, as it happens, for example, in baryon correlators. In fact,
soap films demonstrate a similar behavior to some extent: they easily
form "string junction" with three worldsheets having one common
intersection line and forming e.g. "Mercedes-star" configurations.
Moreover, contrary to the case of one Wilson loop there is no, in
some sense, a unique distinguished string worldsheet configuration
for the product of two or more loops. Let us illustrate this on the
example of two coinciding and oppositely oriented Wilson loops,
where the following well known identity holds true \be | W_{fund}
|^2 = \frac{1}{N_c^2} + \left( 1 - \frac{1}{N_c^2} \right)\cdot
W_{adj} \label{oq2} \ee where $W_{fund}$ is given just by
(\ref{e5}), while $W_{adj}$ corresponds to (\ref{e5}) after
replacement of fundamental $SU(N_c)$ generators $t^a$ with
generators $[T^a]_{bc} = -i f^{abc}$ in adjoint representation and
proper change of trace normalization.\footnote{Traces (both
fundamental and adjoint) are normalized to unity in (\ref{oq2}).}
The terms in the r.h.s. of (\ref{oq2}) are in one-to-one
correspondence with decomposition of the product of two
fundamental $SU(N_c)$ representations into singlet and adjoint:
$N_c \otimes {\bar{N_c}} = {\bf 1} \oplus (N_c^2 - 1)$.
Accordingly, the answer is given by superposition of the state
where two fundamental confining strings are totally annihilated
(corresponding to the unit representation and constant term in
(\ref{oq2})), and the state where confining strings are summed
into one adjoint string (described by the second term in the
r.h.s. of (\ref{oq2})). It is clear that for large contours the
second contribution dies out and the minimal energy state is given
by the singlet state.

Till the present moment we have discussed minimization of
energies, paying no attention to pre-exponential factors like
$1/N_c^2$ in (\ref{oq2}). The latter can become important in
studies of large-$N_c$ behavior of the correlators. We shall come
back to this question later on.

The physical picture outlined above is in fact quite general.
 Therefore we adopt the following algorithm for computing nonperturbative
Wilson loop
correlator. First, for given geometry of the contours, we have to
 find confining string configuration which gives dominant contribution to the
quantity $\lll W(C_1) W(C_2) \rrr$ (notice, that there might be
different surfaces for
 different representations, which appear in the expansion of the
 product of fundamental Wilson loops, as in (\ref{oq2})).
 If the configuration, providing the maximum coincides with
 individual minimal surfaces, we conclude that nonperturbative
 interaction is absent. As the second step, we calculate correlator $\chi^{(0)}(C_1, C_2)$
via (\ref{e10}) in terms of gauge-invariant field correlators.
 We assume that the ensemble of correlators obeys the same hierarchy on this
 surface as it does on the minimal surface for single Wilson loop, i.e.
 exhibits Gaussian dominance. The strategy adopted below
 for computation of higher terms in the expansion (\ref{e9}) is slightly
 different but basic physical arguments for choosing minimal
 surfaces remain the same (see section 4).

We are keeping generality at the moment and perform the vacuum
averaging of Wilson loops, i.e. we explicitly calculate
$\chi^{(0)}(C_1 , C_2)$. Suppose that surfaces $S_1$ , $S_2$
have already been chosen according to our criteria.
 For field-strength tensors, belonging to surfaces $S_1$ and $S_2$
 and gauge-transported to the same point $x_0$ we define:
 $$
 (Fd\sigma)^{(1)}(u) = F_{\mu\nu}(u,x_0) d\sigma_{\mu\nu}(u) \;\; ; \; u
\in
 S_1
 $$
 \be
 (Fd\sigma)^{(2)}(v) = F_{\mu\nu}(v,x_0) d\sigma_{\mu\nu}(v) \;\; ; \; v
\in
 S_2
 \label{e13}
 \ee
 One can now write down the product of two
 Wilson loops in matrix form as
 $$
 N_c^2 \lll W(C_1) W(C_2) \rrr =
 $$
 \be
 = \left\lll \T_1 \T_2 {\mbox P}_{12}
 \left[
 \exp \left(
 ig\int\limits_{S_1} (Fd\sigma)^{(1)}(u)\right)
 \right]_{\alpha_1 \beta_1}
 \left[
 \exp \left(
 ig\int\limits_{S_2} (Fd\sigma)^{(2)}(v)\right)
 \right]_{\alpha_2 \beta_2} \right\rrr
 \label{uu1}
 \ee
 where the traces $\T_1$ and $\T_2$ go over
 indices carrying subscripts $1$ and $2$ respectively
 and the ordering operator ${\mbox P}_{12}$ orders the products of matrices
 in a proper way according to the definition of ${\mbox P}$-exponent
 (\ref{e5}).

One can derive the following rule for vacuum averaging of
several matrix operators, transported to one point
(it is easy to show that these relations are gauge-invariant
under field-independent gauge transformations)
 $$
 \lll [ F(u^{(1)}, x_0) .. F(u^{(k)}, x_0)]_{\alpha_1\beta_1}
 [F(v^{(1)}, x_0) .. F(v^{(m)}, x_0)]_{\alpha_2\beta_2}\rrr =
 $$
 $$
 = \frac{\delta_{\alpha_1 \beta_2} \delta_{\alpha_2 \beta_1} }{
 N_c^2 -1} \left[ \lll \T ( F(u^{(1)}, x_0)..  F(u^{(k)}, x_0)
F(v^{(1)},x_0)
 .. F(v^{(m)} , x_0)) \rrr
 \right.
 $$
 $$
 \left.
 -\frac{1}{N_c} \lll \T ( F(u^{(1)}, x_0)..  F(u^{(k)}, x_0)) \T (
 F(v^{(1)},x_0) .. F(v^{(m)} , x_0))
  \rrr \right] +
 $$
 $$
 \frac{\delta_{\alpha_1 \beta_1} \delta_{\alpha_2 \beta_2}}{
 N_c^2 - 1} \left[
 \lll \T ( F(u^{(1)}, x_0)..  F(u^{(k)}, x_0)) \T ( F(v^{(1)},x_0) ..
 F(v^{(m)} , x_0))\rrr
 \right.
 $$
 \be
 \left.
 -\frac{1}{N_c} \lll \T ( F(u^{(1)}, x_0)..  F(u^{(k)}, x_0) F(v^{(1)},x_0)
 .. F(v^{(m)} , x_0)) \rrr
  \right]
 \label{e171}
 \ee
 For lowest
 Gaussian correlator when $k=m=1$ one has from (\ref{e171})
 \be
 \lll [F(u,x_0)]_{\alpha_1\beta_1}
 [F(v,x_0)]_{\alpha_2\beta_2}\rrr =
 \frac{\lll \T F(u,x_0) F(v,x_0) \rrr }{N_c^2 -1}
 \left( \delta_{\alpha_1 \beta_2} \delta_{\alpha_2 \beta_1} -
 \frac{1}{N_c} \delta_{\alpha_1 \beta_1} \delta_{\alpha_2 \beta_2}
 \right)
 \label{e17}
 \ee
The above relations are valid in theories without global color
symmetry breaking.
 Now one can proceed with matrix cumulant expansion (see, e.g. \cite{vk}):
 \be
 N_c^2 \lll W(C_1) W(C_2) \rrr =
 \T_1 \T_2 {\mbox P}_{12} \exp\left(
 \sum\limits_{n=1}^{\infty} \frac{(ig)^n}{n!} \>
  \llll \tilde F(1)  ...  \tilde F(n) \rrrr \right)
 \label{e14}
 \ee
 where we use the notation
 \be
\tilde F(k) = \int\limits_{S_1} (Fd\sigma)^{(1)}(u^{(k)})
 + \int\limits_{S_2}
(Fd\sigma)^{(2)}(v^{(k)})
 \label{e16}
 \ee
 Here double brackets $\llll ... \rrrr $ denote irreducible correlators
(see
 definition
 in \cite{ddss,vk}).
 The quantity $\tilde F(k)$ carries four independent fundamenatal color indices
and the traces $\T_1 , \T_2$ go over indices
 corresponding to $F^{(1)}$ and $F^{(2)}$, respectively.

Expression (\ref{e14}) provides the basis for our discussion.
For a single loop the corresponding cluster expansion is given by \be
\lll W(C_1) \rrr = \frac{1}{N_c} \>
 \T_1 {\mbox P} \exp\left(
 \sum\limits_{n=1}^{\infty} \frac{(ig)^n}{n!} \>
  \llll \int\limits_{S_1} (Fd\sigma)^{(1)}(u^{(1)}) \cdot ...
\cdot \int\limits_{S_1} (Fd\sigma)^{(1)}(u^{(n)})
 \rrrr \right)
 \label{e144}
 \ee
and in area law regime (which means that typical sizes of the
loops are larger than the gluon correlation length $T_g$ ) one gets
 \be \lll W(C_1) \rrr \propto \exp( -\sigma
S_1^{min}) \label{area} \ee
 The string tension $\sigma$ is given
by \be \sigma S_1^{min} =
\frac{1}{2N_c}\int\limits_{S_1}d\sigma(u^{(1)})
 \int\limits_{S_1}d\sigma(u^{(2)})
  \lll \T gF(u^{(1)},x_0)
 gF(u^{(2)},x_0) \rrr + ...
\label{stp8}
\ee
where the dots denote higher non-Gaussian terms.
As it was mentioned above, we do not take into account perimeter
terms, it is implicitly supposed that
all considered loops are large enough in this sense.
 It is straightforward to rewrite
(\ref{e14}) combined with (\ref{e144}) and (\ref{e171}) in the
following way
 \be
 N_c^2 \lll W(C_1) W(C_2) \rrr =
 \T_1 \T_2 \exp \left(\hat 1 \cdot(\Lambda_0 +
 \Lambda_1) + {\hat e}\cdot\Lambda_e \right)
 \label{e18}
 \ee
where \be \Lambda_0 = -\sigma S_{1} - \sigma S_2 \label{tr} \ee
and terms $\Lambda_1$ and $\Lambda_e$ contain correlators of
powers of $F$ defined on different surfaces and hence provide
contribution to $\chi^{(0)}$. They are as follows:
 \be
 \Lambda_1 =
 \pm \frac{g^2}{N_c(N_c^2 -1)} \int\limits_{S_1}d\sigma(u^{(1)})
 \int\limits_{S_2}d\sigma(v^{(1)})
  \lll \T F(u^{(1)},x_0)
 F(v^{(1)},x_0) \rrr + ...
 \label{e19}
 \ee
 and
 \be
 \Lambda_e = \mp
 \frac{g^2}{N_c^2 -1} \int\limits_{S_1}d\sigma(u^{(1)})
 \int\limits_{S_2}d\sigma(v^{(1)})
  \lll \T F(u^{(1)},x_0)
 F(v^{(1)},x_0) \rrr + ...
 \label{e20}
 \ee
 where the dots again denote higher non-Gaussian terms.
 The upper (lower) sign in the above expressions corresponds to the case of
 parallel (opposite) orientation of the contours $C_1$ and $C_2$ and surfaces
$S_1$ and $S_2$
(we assume that the orientation of the surface is fixed by the orientation of
its boundary contour).

The matrix structures $\hat 1$ and $\hat e$ introduced above are given in
 index
 notation by
 \be
 [\hat 1]_{\alpha_1\beta_1;
 \alpha_2\beta_2} = \delta_{\alpha_1 \beta_1} \delta_{\alpha_2 \beta_2}
 \;\; ; \;\;
 [\hat e]_{\alpha_1\beta_1;
 \alpha_2\beta_2} = \delta_{\alpha_1 \beta_2} \delta_{\alpha_2 \beta_1}
 \label{e21}
 \ee
 We are to examine what algebra the matrices
 $\hat 1$ and $\hat e$ do obey. In the chosen approximation it encodes the
 effect of ordering, performed by the operator ${\mbox P}_{12}$.
 Consider the case of parallel orientation. In this case
 ordering of the fields $F^{(1,2)}$ in the Wilson loop product coincide for
 both loops, which for the matrices $\hat 1$ and $\hat e$ mean,
 that they should be multiplied from the right with respect to indices
 carrying subscript "1" and corresponding to the surface $S_1$, and
 also with respect to indices
 carrying subscript "2" and corresponding to the surface $S_2$.
  It is easy to check, that this requirement transforms into the
 following relations:
 \be
 \hat e \cdot \hat 1 = \hat 1 \cdot \hat e = \hat e \; ; \; {\hat 1}^2 =
\hat
 1
 \; ; \; {\hat e}^2 = \hat 1
 \label{e25}
 \ee

 In case  of antiparallel orientations of surfaces the matrix $\hat e $
 should be multiplied from the right with respect to indices
 carrying subscript "2" but
 from the left with respect to indices
 carrying subscript "1", which corresponds to the fact of opposite
 ordering of the fields $F^{(1)}$ and $F^{(2)}$. It results in the algebra
 different from (\ref{e25}):
 \be
 \hat e \cdot \hat 1 = \hat 1 \cdot \hat e = \hat e \; ; \; {\hat 1}^2 =
\hat
 1
 \; ; \; {\hat e}^2 = N_c \hat e
 \label{e26}
 \ee
 We also notice that in both cases $\T_1 \T_2 \> \hat 1 = N_c^2$ and
 $\T_1 \T_2 \> \hat e = N_c$.

 With (\ref{e25}), (\ref{e26}) at hand, we can easily compute
 (\ref{e18}). For parallel orientation of surfaces one gets
$$
 \chi^{(0)}(C_1, C_2) =
 \left[ \frac12 \left(1-\frac{1}{N_c}\right)\exp\left(\Lambda_0 +
 \Lambda_1 - \Lambda_e \right) + \right.
$$
\be + \left.
  \frac12 \left(1+\frac{1}{N_c}\right)\exp\left(
\Lambda_0 +
 \Lambda_1 + \Lambda_e
  \right) \right]
- \exp\left(\Lambda_0^{min} \right)
 \label{e273}
 \ee
 where the last term  corresponds to the product of averages
 of two loops.
  For oppositely directed contours the result is
 \be
 \chi^{(0)}(C_1, C_2) =   \left[
 \frac{1}{N_c^2} \exp\left( \Lambda_0 +
 \Lambda_1 + N_c \Lambda_e \right) \> +
 \left(1 -
 \frac{1}{N_c^2}\right)
 \exp\left(\Lambda_0 + \Lambda_1 \right)\right]
-\exp\left( \Lambda_0^{min}\right)
 \label{e283}
 \ee

Expressions (\ref{e273}), (\ref{e283}) together with the
prescription for the choice of optimal integration surface provide
the answer for nonperturbative interaction term. Let us come to
concrete examples and first consider the simplest possible case
when $S_1 = S_1^{min} , S_2 = S_2^{min}$ and $S_2^{min} \subset
S_1^{min}$ (see Fig.1). As is was explained in details above, we
always assume Gaussian dominance.
 One easily gets the following result for the contours with linear sizes
greater than $T_g$ and omitting perimeter contributions:
 \be
 \Lambda_1 = \pm \frac{2}{N_c^2 -1}
 \sigma S_2^{min} \;\;\; ; \;\;\;
 \Lambda_e = \mp
  \frac{2N_c}{N_c^2 -1}
 \sigma S_2^{min}
 \label{e23}
 \ee
and (\ref{e273}), (\ref{e283}) become
 $$
 \chi^{(0)}(C_1, C_2) = \exp(-\sigma S_1^{min} - \sigma S_2^{min} )
 \cdot
 \left[ \frac12 \left(1-\frac{1}{N_c}\right)\exp\left(\frac{2\sigma
 S_2^{min}}{N_c -1}\right) + \right.
$$
\be + \left.
  \frac12 \left(1+\frac{1}{N_c}\right)\exp\left(- \frac{2\sigma S_2^{min}}{N_c
 +1}\right) \right]
- \exp(-\sigma S^{min}_1 - \sigma S^{min}_2).
 \label{e27}
 \ee
 for parallel orientations and
 $$
 \chi^{(0)}(C_1, C_2) =  \exp(-\sigma S_1^{min} - \sigma S_2^{min} )\cdot \left[
 \frac{1}{N_c^2} \exp\left( 2\sigma S_2^{min}\right) \> + \right.
$$
\be + \left. \left(1 -
 \frac{1}{N_c^2}\right)
 \exp\left(-\frac{2\sigma S_2^{min}}{N_c^2 -1}\right)\right]
-\exp(-\sigma S^{min}_1
-\sigma S^{min}_2).
 \label{e28}
 \ee
 for opposite orientations.
 It is worth reminding that $S_1^{min} > S_2^{min}$ by our convention.

Notice that in case of
 coinciding, but oppositely directed contours (i.e. $C = C_1 =
 [C_2]^{\dagger}$),
 expression (\ref{e28}) reproduces (\ref{oq2}), if adjoint string tension
is given by the Casimir ratio
 \be
 \sigma_{adj} = \frac{2 N_c^2}{N_c^2 - 1} \sigma
 \label{e29}
 \ee
This result is to be expected since, as it was already mentioned,
Gaussian dominance yields Casimir scaling \cite{casimir}. If  both
contours $C_1,C_2$
 lie on the same plane and $C_2$ is inside $C_1$,
the geometry becomes effectively two dimensional and the results
(\ref{e27}), (\ref{e28}) coincide with
 formulas obtained in slightly different way in \cite{dk}.
The same expressions
 hold true in 1+1 dimensional Yang-Mills theory, where
one has just two dimensional geometry and, on the other hand,
exact Gaussian picture. One also look upon (\ref{e27}),
(\ref{e28}) as an algebraic rule of adding up parallel or
antiparallel fundamental fluxes which illustrate the decomposition
${\bf 3} \otimes {\bf 3}={\bf \bar{3}} \oplus {\bf 6}$ and ${\bf
3} \otimes {\bf \bar{3}}={\bf 1} \oplus {\bf 8}$ respectively,
with the string tension
 in each representation given by the Casimir scaling law .

We are now interested in the case of contours separated by
distances greater than $T_g$. By way of example let us calculate
purely nonperturbative correlator of two Wilson loops $\lll W(C_1)
W(C_2) \rrr $ for simple rectangular geometry of the contours. We
choose two rectangular contours $R\times T$ lying on parallel
planes, at distance $h$ from each other (see Figs. 2, 3). We
suppose that $T \gg R$ and will not take care of subleading $1/T$
terms. If $h$ is of the order of $T_g$, one comes back to the case
described by (\ref{e27}), (\ref{e28}). We take the distance $h$
such that $R \gsim h \gsim T_g$, where the nonperturbative regime
is supposed to play a role. One could still choose for $S_1$ and
$S_2$ the corresponding minimal surfaces $S_1^{min}$ and $
S_2^{min}$. In this case the correlator $\chi^{(0)}$ would be
equal to zero up to exponentially small terms of the kind
$\exp(-2h/T_g)$, as it is clearly seen from (\ref{e19}),
(\ref{e20}). Such configuration of surfaces however does not
correspond to the minimal energy condition formulated above and is
therefore unstable. It is clear that the subtraction term, the
last exponent
 on the r.h.s. of (\ref{e27}), (\ref{e28}) stays intact, since it is a product
 of single loop averages. Other terms, on the contrary, strongly
 depend on the profile of the strings, which defines the
 integration surface and is to be chosen according to dynamical minimal
 energy condition.
Consider first the case of opposite loop orientations and let us
examine different choices of surfaces. As a trivial example we
might adopt the same choice as above, namely $S_1 = S_1^{min} ;
S_2 =S_2^{min}$. The first term in square brackets in (\ref{e283})
contributes to $\lll W(C_1) W(C_2) \rrr $ as
$$
\frac{1}{N_c^2} \exp\left( -2\sigma RT \right)
$$
as it should be since $\Lambda_0 = \Lambda_0^{min} = \sigma
S_1^{min} + \sigma S_2^{min} = 2\sigma RT$ and $\Lambda_1 =
\Lambda_e = 0$ for this choice of surfaces (up to exponentially
small terms $\sim \exp(-2h/T_g)$, which we always omit in the
paper). Correspondingly, the second term in square brackets in
(\ref{e283}) would be
$$
\left( 1 - \frac{1}{N_c^2}\right) \exp\left( -2\sigma RT \right)
$$
with the sum of two giving expected answer $ \lll W(C_1) \rrr \lll
W(C_2) \rrr = \exp(-2\sigma RT) $ and hence $\chi^{(0)} = 0$. One
immediately sees that such choice is not optimal if $h$ is small.
Instead, if we choose $S_1$ as a
 minimal enveloping surface with boundary on $C_1$ and coinciding with
 $S_2 = S_2^{min}$ inside $C_2$ (due to apparent symmetry of our problem,
 one could of course easily interchange indices '1' and '2'), we get
for the first term in square brackets in (\ref{e283})
$$
\frac{1}{N_c^2} \exp\left( -2\sigma hT \right)
$$
where $2hT = S_{12} = S_1 - S_2$. The second term contributes to
$\lll W(C_1) W(C_2) \rrr $ as
$$
\left( 1 - \frac{1}{N_c^2}\right) \exp\left( -2\sigma hT -
\sigma_{adj} RT\right)
$$
where $\sigma_{adj}$ is given by (\ref{e29}) and this contribution
is always subleading with respect to the former one if $R \gg h$
and $N_c$ is not exponentially large (see below). So these two
different choices give different answers for Wilson loop
correlator: \be \lll W(C_1) W(C_2) \rrr =  \exp(-2\sigma RT)
\label{eu} \ee in the first case (two individual minimal surfaces)
and \be \lll W(C_1) W(C_2) \rrr = \frac{1}{N_c^2} \exp\left(
-2\sigma hT \right) + \left( 1 - \frac{1}{N_c^2}\right) \exp\left(
-2\sigma hT - \sigma_{adj} RT\right) \label{eu3} \ee in the second
case (enveloping geometry). According to our criteria, the answer
which is dominant should be chosen as correct, since it
corresponds to actual string configuration. It is seen, that there
is a critical distance between loops \be h_{crit} \approx R -
\frac{1}{\sigma T} \log N_c \label{en} \ee
 in our problem.\footnote{For abelian confining strings
 including soap films the last term in (\ref{en}) is absent.}
 For $h < h_{crit}$ confining strings rearrange
 themselves\footnote{See also \cite{dis}, where static multiquark
 interactions were studied in strong coupling expansion regime.}
 with respect to noninteracting case,
 which is encoded in expression (\ref{eu3}). Correspondingly,
 one has nonzero $\chi^{(0)}$. For larger
$h$ they do not interact and $\chi^{(0)}(C_1 , C_2)$ vanishes. It
is important that in our picture it happens dynamically, in
particular, one cannot just naively take large $N_c$-limit in
(\ref{e273}), (\ref{e283}). On the other hand, if $h$ is kept
fixed, then it is clear from (\ref{en}) that in large $N_c$ limit
$\chi^{(0)}$ should vanish. It is worth saying a few words about
the meaning of $T$ in (\ref{eu}), (\ref{eu3}). It is seen that
with $T$ going to infinity, $h_{crit}$ is increasing and
approaching $R$. Therefore for long static loops minimal energy state
will always eventually win, as it should be. If however one
studies not correlators of static loops but some physical process
like hadron scattering, pre-exponent factors are important and $T$
corresponds to the typical interaction time. In particular cases
it can be quite small. Expression (\ref{en}) then physically means
that if $T$ is small, strings does not have enough time to
rearrange and nonperturbative interaction does not take place.

To summarize, the answer is given by \be \chi^{(0)} =
\frac{1}{N_c^2} \exp\left( -2\sigma hT \right) + \left( 1 -
\frac{1}{N_c^2}\right) \exp\left( -2\sigma hT - \sigma_{adj}
RT\right) -  \exp\left( -2\sigma RT\right) \label{eiu} \ee if
$h<h_{crit}$ and $\chi^{(0)} = 0$ if $h>h_{crit}$. We now come to
the case of parallel orientations. Let us consider two geometries
analyzed above. The case of individual minimal surfaces is
identical for parallel and opposite orientations and has just been
considered. The enveloping geometry gives for the first term in
square brackets in (\ref{e273})
$$
 \frac12 \left(1-\frac{1}{N_c}\right)\exp\left(-2\sigma hT -
\frac{2\sigma RT(N_c -2)}{N_c -1}\right)
$$
while the second term becomes
$$
\frac12 \left(1+\frac{1}{N_c}\right)\exp\left(-2\sigma hT -
\frac{2\sigma RT(N_c +2)}{N_c +1}\right)
$$
and is always subleading. In particular case $N_c = 3$ the former
term corresponds to the creation of the string in ${\bf \bar 3}$
representation of the length $R$ which has the same tension as the
fundamental string while the second term describes sextet
representation string formation ${\bf 6} = {\mbox{Symm}}\{ {\bf 3}
\otimes {\bf 3}\}$ with tension $2(N_c+2)/(N_c + 1)=2.5$ times
larger than fundamental one. In this particular case of $N_c = 3$
one can additionally consider double Y-shape profile, shown on
Fig.3, with the result for the first term in square brackets in
(\ref{e273})
$$
\frac{1}{3}\exp(- \sqrt{3}\sigma h T - \sigma RT)
$$
Since $\sqrt{3} < 2$, this term is seen to be dominant over
enveloping geometry in $N_c =3$ case.

As in opposite orientations case there appears critical length
$h_{crit}$ which for $N_c =3$ is given by
 \be h_{crit}=\frac{1}{\sqrt{3}}\left( R - \frac{\log
3}{\sigma T} \right)
 \ee
The expression for $\chi^{(0)}$ becomes ($N_c = 3$):
$$
 \chi^{(0)}(C_1, C_2) = \exp(-\sigma S_{12} )
 \cdot
 \left[ \frac13 \exp\left(-\sigma
 S_3\right) + \right.
$$
\be + \left.
  \frac23 \exp\left(- \frac{5\sigma S_3}{2}\right) \right]
- \exp(-2\sigma RT)
 \label{ew29}
 \ee
where $S_{12}$ represents the boundary surface $S_{12} =
4hT/\sqrt{3} $ and $S_3 = (R-h/\sqrt{3})T$ is the common part of
$S_1$ and $S_2$. Notice that $l_{crit} = R - h_{crit}/\sqrt{3} >
0$, therefore one is never in the situation of $S_3$ shrinking to
zero. In some sense $l_{crit}$ in (\ref{ew29}) plays the role of
$R_{min}$ from (\ref{er5}).

 The physical picture is the same as for opposite orientations -- at small
distances $h$,  $h_{crit} > h \gsim T_g$ preferable string
configuration is given by double Y-shape profile shown on Fig.3
while at larger distances there is no common string state formation and
hence nonperturbative interaction is absent, $\chi^{(0)}(C_1 ,
C_2) = 0$.

It is important to compare at this point our approach to  another
approach, which is also based on field correlator method, but
follows another logic. This well-known strategy started in
\cite{doschn} has many successful phenomenological applications
and modifications (see, e.g. \cite{appl}).
 There are two essential differences
to be mentioned. The first one comes from the fact that we are
working in Euclidean metric rather than in Minkowski. It is
important to stress, that in our approach the vacuum averaging
over gluon fields (which eventually creates the string) is to be
done in Euclidean space with Euclidean gluon configurations, and
transition to Minkowski space can be done only after this vacuum
averaging. This is intimately connected to the realization that
vacuum configurations are of tunnelling type and it is not
legitimate to continue them analytically into Minkowski space.
However, as soon as the fields are integrated out the transition
to the latter is not a problem and as we demonstrated in
\cite{dks}  the QCD string is present also on the light cone.
 Correspondingly, we do not have a notion
of light-like surfaces with vanishing areas, which is of
importance in the formalism developed in \cite{doschn}. The second
point is a different physical picture behind the purely
nonperturbative interaction in our model. The magnitude of
$\chi^{(0)}(C_1, C_2)$ depends in our approach on the chosen
string worldsheet configuration which, in its turn, is determined
by the dynamical minimal energy requirement, discussed above.
However, we do not study effects of string overlap, caused by
finite thickness of the strings. If the minimal surface
requirement dictates $S_{1, 2} = S_{1, 2}^{min}$, which always
happen for well separated loops, there is no purely
nonperturbative interaction in our picture, i.e. $\chi^{(0)}(C_1,
C_2) = 0$. This is in sharp contrast to the approach \cite{doschn}
where all physical effect is due to nonzero value of $T_g$ which
bring about nonzero overlap between hadron Wilson loops. In other
words, the expressions (\ref{e27}), (\ref{e28}) have nontrivial
limit if $T_g \to 0$, while string tension $\sigma \sim \lll F^2
\rrr T_g^2$ is kept fixed, contrary to the corresponding
expressions from \cite{doschn,appl} which vanish in this limit.
Accordingly the hadron--hadron cross-section in that approach is
proportional to $T_g$ in the tenth power \cite{appl} and is very
sensitive to exact value of $T_g$. As it is, both approaches are
viable in their regions of values of $T_g$, and it is very
important to extract this value from lattice data, analytically or
phenomenologically. The existing calculations yield values in the
range $0.35$ Fm $> T_g > 0.1$ Fm for different lattice and
analytic procedures (see \cite{dig} and references therein). We
assume in our approach that $T_g$ is closer to lower limit and
therefore the string overlap effect proportional to higher powers
of $T_g$ is a small correction to our basic mechanism. One should
stress however, that if even $T_g$ would be large to justify
approach of \cite{doschn,appl} at moderate energies, the
gluon-exchange--generated Pomeron mechanism is always dominating
asymptotically at large energies.

To conclude this section let us say a few words about the case of
distant loops.
 The general expression (\ref{e14}) is applicable here as well.
 In Gaussian approximation one still has (\ref{e18}). Important change
 however
 is that now correlator $\lll F^{(1)} F^{(2)} \rrr $ does not  get
any contributions since the surfaces $S_1, S_2$ should be chosen
according to our principle, as independent and not intersecting
minimal surfaces. Hence nonperturbative contribution vanishes in
this case, up to the string overlap effects mentioned above.

 In other words when contours are
 distant from each other, it is always preferable to deform $S_{12}$ to two
 discs, corresponding to minimal surfaces for
 each contour plus whatever thin tube connecting these surfaces through the
 point
 $x_0$, thus reducing $\chi^{(0)}(C_1, C_2)$ to zero.
 This is a sign, that purely nonperturbative contribution vanishes,
 and
 one should consider next terms in the expansion (\ref{e9}), namely
 $\chi^{(4)}(C_1, C_2)$. It is important to realize that the surface
entering
 (\ref{e10}) is not dynamical\footnote{It has come from the Stokes theorem
 and therefore could be arbitrary, subject to our principle of minimal action.},
 which reflects itself
 in the possibility of infinite squeezing of the tube, connecting two
distant
 minimal surfaces. The situation changes however when one includes
 perturbative
 gluons, propagating inside its wall and forming a physical glueball state
in
 this way. It is actually this glueball exchange mechanism, which
corresponds
 to the term  $\chi^{(4)}(C_1, C_2)$ when background field is taken into
 account.
 We study this term in the next secton.

 \section{Glueball exchange interaction}

 We now turn to the term $\chi^{(4)}(C_1, C_2)$ in the expansion
 (\ref{e9}). We shall see that in the absence of NP background this
 term reduces to the purely perturbative two-gluon exchange term
 suggested in \cite{low} as a basic element of Pomeron exchange.
 To be more precise, let us consider first the case of no background
fields.
 In the Feynman gauge for $a_{\mu}$, one has in the lowest
 order
 \be
 \lll a_{\mu}^a(x) a_{\nu}^b(y) \rrr = \dl_{\mu\nu} \dl^{ab} \>
 \frac{1}{4\pi^2 (x-y)^2}
 \ee
 Our expression for $\chi^{(4)}$ will have the same form
 as (\ref{e18}), where one should keep only
 nonperturbative background field $B_{\mu}$ in diagonal correlators
 $\lll \T F^{1} F^{1} \rrr$ while terms proportional
 to $\lll \T F^{1} F^{2} \rrr$ contain only perturbative exchange.
 These perturbative exchanges are modified however by the presence
 of nonperturbative background. To take it into account, one has to perform
 averaging
 in two steps: first in valence (perturbative) field $a_{\mu}$ and
 second - in background field $B_{\mu}$:
 $$
 \lll W(C_1) W(C_2) \rrr = \frac{g^4}{N_c^2}
 \T_1\T_2 {\mbox P}_{12} \int d x_{\mu_1}^{(1)} \int
 d x_{\mu_2}^{(2)} \int d y_{\nu_1}^{(1)}\int d y_{\nu_2}^{(2)}
 $$
 $$
 \Phi_{C_1}(x^{(2)} , x^{(1)}) t^{a_1}
 \Phi_{C_1}(x^{(1)} , x^{(2)}) t^{a_2}
 \Phi_{C_2}(y^{(2)} , y^{(1)}) t^{b_1}
 \Phi_{C_2}(y^{(1)} , y^{(2)}) t^{b_1} \cdot
 $$
 \be
 \cdot ( \lll a(x^{(1)}) a(y^{(1)}) \rrr
 \lll a(x^{(2)}) a(y^{(2)}) \rrr + \lll a(x^{(1)}) a(y^{(2)}) \rrr \lll
 a(x^{(2)}) a(y^{(1)}) \rrr )
 \label{eo12}
 \ee
 where short-hand notation was used $a(x^{(1)}) \equiv
 a_{\mu_1}^{a_1}(x^{(1)})$. The coordinates $x^{(1)}$, $x^{(2)}$ are
 ordered along the contour $C_1$ as well as
 $y^{(1)}$, $y^{(2)}$ are
 ordered along the contour $C_2$.
 Notice also gauge-invariance of (\ref{eo12}) due to transformation law
 (\ref{e2}).

 Before proceeding further
 one is to define the dependence of $\lll a_{\mu}(x) a_{\nu}(y) \rrr$
 on background fields. To this end it is convenient to use
Feynman-Schwinger
 representation for gluon Green's function \cite{ffs} and represent it as
 \be
 G_{\mu\nu}^{ab} = \lll a_{\mu}^a(x) a_{\nu}^b(y) \rrr =
 \int\limits_0^{\infty} ds \left({\cal D} z \right)_{xy}
 \exp (-K_0) \Phi_{\mu\nu}^{ab}(x,y)
 \label{e199}
 \ee
 where
 $$
 K_{0} = \frac14 \int\limits_0^{\infty} d\tau \left(
 \frac{d z_{\mu}}{d\tau}\right)^2
 $$
 and
 $$
 \Phi_{\mu\nu}^{ab}(x,y) = \left[
 \mbox{P}_F \mbox{P}_A \exp \left(ig \int\limits_y^x A_{\mu}(z)
 dz_{\mu}\right)
 \exp\left(2g \int_0^s d\tau F(z(\tau))\right) \right]_{\mu\nu}^{ab}
 $$
 and $a, b$ are adjoint color indices, whereas $\mu , \nu$ - Lorentz
 indices, i.e. $[F]_{\mu\nu}^{ab} = -i F_{\mu\nu}^c f^{abc}$.

 To understand better the topology of the resulting construction,
 it is useful to consider large $N_c$ limit. One can write in this limit
for
 adjoint
 phase factors in (\ref{e199})
 $$
 [ t^a]_{\alpha\beta} \Phi^{ab}(x,y)
 [ t^b]_{\gamma\delta} =
 {\mbox P}_{A} \exp \left( ig \int\limits_y^x {\hat A}_{\lambda}
dz_{\lambda}
 \right)^{ab}   [ t^a]_{\alpha\beta} [ t^b]_{\gamma\delta} =
 $$
 \be
 = \Phi_{\alpha\delta}(x,y) \Phi_{\gamma\beta}(y,x) + {\cal O} \left(
 \frac{1}{N_c} \right)
 \label{e1998}
 \ee
 where $\Phi_{\alpha\delta}(x,y)$ is parallel transporter in fundamental
 representation.
 Expression (\ref{e1998}) exemplifies well known 't Hooft's rule for
replacing
 gluon line by double adjoint in large-$N_c$ limit. Inserting (\ref{e1998})
into (\ref{eo12}),
 one obviously obtains two new Wilson loops $C_{12}$ and ${C_{12}}'$
instead
 of previous $C_1$
 and $C_2$: each initial loop is now divided by two gluon
 emissions/absorptions
 into two arcs which are connected by double lines of gluon propagators (see
 Figs. 4,5).
 At small $N_c$ this construction goes over into that of two fundamental
 loops
 $C_1$ and $C_2$ connected by two adjoint lines and final result will
amount
 to replacing double fundamental string worldsheet by one adjoint string
 wordsheet.
 In terms of string tensions it correspond to replacement of
 $2\sigma$ by $9\sigma / 4$ in Gaussian approximation.
 We will keep large $N_c$ limit and replacement (\ref{e1998})
 in what follows.

 The averaging over background fields leads to the following result:
 $$
 \lll\lll W(C_1) W(C_2) \rrr_{a}^{(4)} \rrr_B = \chi ^{(4)}(C_1, C_2) =
 $$
 \be
 = \frac{{\tilde g}^4}{N_c^2} \int\limits_{L_{12}} ds_1 {\cal D}
 z^{(1)} \int\limits_{{L_{12}}'} ds_2 {\cal D} z^{(2)} \>
 \exp(-K_0^{(1)} - K_0^{(2)} )\> \lll W(C_{12}) W({C_{12}}') \rrr
 \label{eew}
 \ee
 where ${\tilde g}^2 = g^2 N_c$.
 Here contours $C_{12}$ and ${C_{12}}'$ comprise pieces of $C_1$
 and $C_2$ connected by two double fundamental lines
 $L_{12}$ and ${L_{12}}'$. It is understood that surfaces $S_{12}$,
 ${S_{12}}'$ are subjects of our general assumption about minimal
 action. This gives different forms depending on the distance
 between original loops $C_1$ and $C_2$, see below.
 It is also understood that gluon spin operators $2g F(z)$
 are to be placed on the gluon trajectories $L_{12}$ and ${L_{12}}'$
 in accordance with (\ref{e199}). It will produce gluon
 spin interaction terms which influence glueball Green's function;
 to simplify discussion we omit these terms at the moment.

 Now we can use large-$N_c$
 factorization property for the product $\lll W(C_{12}) W({C_{12}}') \rrr $
 and use area law asymptotics for each piece, i.e.
 for surfaces $S_{12}$ and ${S_{12}}'$. One obtains
 \be
 \chi ^{(4)}(C_1, C_2) =
\frac{{\tilde g}^4}{N_c^2} \int\limits_{L_{12}} ds_1 {\cal D}
 z^{(1)} \int\limits_{{L_{12}}'} ds_2 {\cal D} z^{(2)} \>
 \exp(-K_0^{(1)} - K_0^{(2)} )\> \exp(-\sigma (S_{12} + {S_{12}}'))
 \label{e1ew}
 \ee

 To define the profiles of the surfaces we shall use the same
 principle outlined above, i.e. we require an  effective value of
 area $\lll S_{12} + {S_{12}}' \rrr$ averaged over possible gluon
 trajectories
 $L_{12}$ and ${L_{12}}'$ to be minimal.
 The result will of course strongly depend on relative positions and
 orientations of the contours $C_1$ and $C_2$.
 In the first case when both loops lie on the same plane and $C_2$ is
 entirely inside $C_1$, it is clear that the sum $S_{12} + {S_{12}}' $
 does not depend on trajectories $L_{12}$ and ${L_{12}}'$, and one has
 $S_{12} + {S_{12}}' = S_1 - S_2$. Thus one obtains
 effectively the surface $S_1$ with the hole due to $C_2$, i.e. a
construction
 which already has appeared in purely nonperturbative term for opposite
 oriented contours $C_1, C_2$, but now with two valence gluons connecting
 contours $C_1$ and $C_2$, see Fig.4.

 However for large enough distances the true minimum of
 $  S_{12} + {S_{12}}' $ is reached by another construction --
 when two contours $C_1$ and $C_2$ are connected by narrow
 strip formed by trajectories $L_{12}$ and ${L_{12}}'$
 with the double (adjoint) string worldsheet between them (Fig.5).
 This narrow strip is nothing but the glueball Green's
 function and the width of the strip
 is equal to the average size of the lowest mass glueball,
 i.e. around $0.5$ Fm. Notice that due to kinetic terms
 in (\ref{e1ew}) this strip is dynamical (contrary to the nonperturbative
 case) and cannot be shrinked.  This is a typical construction for
 high-energy scattering amplitude when the glueball exchange diagram
 is gradually replaced by glueball Regge trajectory exchange, i.e.
 by Pomeron exchange, which persists to larger
 experimentally accessible energies.

 To demonstrate that explicitly, one should rewrite expression
(\ref{e1ew})
 directly in terms of glueball Green's function:
 $$
 \frac{\chi ^{(4)}(C_1, C_2)}{\lll W(C_1) W(C_2) \rrr } =
 $$
 \be
 {\mbox P}_{12} \int\limits_{C_1} dx_{\mu_1}^{(1)}
 \int\limits_{C_1} dx_{\mu_2}^{(2)}
 \int\limits_{C_2} dy_{\nu_1}^{(1)}
 \int\limits_{C_2} dy_{\nu_2}^{(2)} \left[
 G_{\mu_1 \mu_2}^{ \nu_1 \nu_2 } (x^{(1)}, x^{(2)} | y^{(1)} , y^{(2)} ) +
 ( y^{(1)} \leftrightarrow y^{(2)} ) \right]
 \label{yy}
 \ee
 where $G_{\mu_1 \mu_2}^{ \nu_1 \nu_2 } (x^{(1)}, x^{(2)} | y^{(1)} ,
 y^{(2)} ) $ is
 two-gluon glueball Green's function, describing propagation
 from points $x^{(1)}, x^{(2)} $ to $ y^{(1)} , y^{(2)}$, which has
 the Feynman-Schwinger representation as in (\ref{e1ew}). The spectrum
 of this Green's function (with spin terms included) was computed
 analytically in \cite{kaid}.

 In case all points are close to each other, i.e.
 $$
 | x^{(i)} - x^{(j)} | \ll T_g \;\; ; \;\;  | x^{(i)}  - y^{(j)} | \ll T_g
 $$
 one can replace glueball Green's function by a product of
 free gluon propagators
 \be
 G_{\mu_1 \mu_2}^{ \nu_1 \nu_2 } (x^{(1)}, x^{(2)} | y^{(1)} , y^{(2)} )
\sim
 g^4 \frac{\dl_{\mu_1 \nu_1} \dl_{\mu_2 \nu_2}}{(2\pi)^4 (x^{(1)} -
 y^{(1)})^2
 (x^{(2)} - y^{(2)})^2 }
 \label{free}
 \ee
 Another asymptotics is available when both
 $| x^{(i)}  - y^{(j)} |$ are large, then the spectral decomposition is
 possible:
 $$
 G_{\mu_1 \mu_2}^{ \nu_1 \nu_2 } (x^{(1)}, x^{(2)} | y^{(1)} , y^{(2)} )
\sim
 $$
 \be
 \sim
 \sum\limits_n \Psi_{\mu_1 \mu_2}^{(n)} (x^{(1)} , x^{(2)})
 {\Psi_{\nu_1 \nu_2}^{(n)}}^{\dagger} (y^{(1)} , y^{(2)})
 \cdot \exp \left(
 - M_{n} \cdot \left| \frac{x^{(1)} + x^{(2)}}{2} -  \frac{ y^{(1)} +
 y^{(2)}}{2} \right| \right)
 \label{f1ree}
 \ee

 Since the lowest glueball is rather heavy, $M_0 \approx 1.5$ GeV, one
expect
 fast decrease of $\chi^{(4)}$ when distance between loops is growing:
 $$
 \chi^{(4)}(h) \sim \exp(-M_0 |h|)
 $$
 The situation is qualitatively similar to the one studied in
 \cite{mul1,dip}, where gluon was assumed to have effective mass
 $m_g \sim 0.9$ GeV. One expects dipole-dipole cross section
 around a few mb in this case, when $\alpha_s$ is of the order of one.
To obtain realistic large hadron-hadron scattering one
 needs glueball exchange to be reggeized, in which case radius
 of interaction grows logarithmically   \cite{gribov}.
 In particular case of BFKL Pomeron this picture was studied
 in \cite{bfkl}. Our picture differs from that of BFKL, since nonperturbative
 background is taken into account.
Let us consider as an example the problem of high-energy forward
onium-onium scattering. Since the interaction time between
particles at high energies is much smaller than the typical
interaction time for quarks inside onium, one can consider the
onium in this process as free quark-antiquark pair (see, e.g.
\cite{mul1}). The small radius of the onium compared with typical
transversal length scales of the problem dictates
$\chi^{(4)}$--dominance over $\chi^{(0)}$ in the problem since
possible nonperturbative string configurations which could
contribute to $\chi^{(0)}$ are strongly suppressed over individual
minimal noninteracting strings.

Since it is more convenient to study the scattering of systems in
given quantum states rather than scattering of Wilson loops, we
are to switch from (\ref{g4w}) to spectral decomposition of the
form (\ref{sur}) and take only one term, corresponding to the
scattering of particular states. The resulting expression
coincides (up to normalization factor) with the scattering
amplitude (see, e.g. \cite{peskin}). Since in the Feynman gauge
for the field $a_{\mu}^a$ we have \be G_{\mu_1 \mu_2}^{ \nu_1
\nu_2 } (x^{(1)}, x^{(2)} | y^{(1)} , y^{(2)} ) = \dl_{\mu_1
\nu_1} \dl_{\mu_2 \nu_2} G^{(2)} (x^{(1)}, x^{(2)} | y^{(1)} ,
y^{(2)} ) \label{es4} \ee the answer can be straightforwardly
obtained using the same strategy as in \cite{mul1}: \be T_{forw} =
i \int d^2 \rho_1 \int d^2 \rho_2 \int\limits_0^1 dz_1
\int\limits_0^1 dz_2
 |\psi(\rho_1, z_1)|^2 |\psi(\rho_2, z_2)|^2 F(\rho_1, \rho_2, Q=0) \label{ff} \ee
 where
 $$
 F(\rho_1 , \rho_2 , Q=0) = \frac{N_c^2 -1}{32 \pi^2 N_c^2} \>
 \int d^2 k \> G^{(2)}(k, Q = 0) \cdot
$$
\be
 \cdot (2-\exp(-ik\rho_1) - \exp(ik\rho_1))(2-\exp(-ik\rho_2) -
 \exp(ik\rho_2))
\label{opia} \ee In the above expression $G^{(2)}(k, Q) $ is
Fourier transform of (\ref{es4}) with respect to total momentum
$p_1 + p_2 = Q$ and relative momentum $p_1 - p_2 = Q-2k$, the
former is equal to zero for forward scattering amplitude (we also
suppose in (\ref{opia}) vanishing transverse momenta of onia). The
mixed representation wave function $\psi(\rho_1, z_1)$ defined on
the light-cone describes the state of color dipole with transverse
size $\rho_1$ and fraction of total onium light-cone momentum
$p_1^{+}$ carried by quark $z_1$. We omit spinor indices, assuming
proper summation over them. If one will "turn off" confinement
(i.e. in our formalism put the confining background field to zero
everywhere), the Green's function in the leading order of
perturbation theory would be just a product of two gluon
propagators \be G^{(2)}_{free}(k, Q)= \frac{g^4}{(Q-k)^2 k^2}
\label{asq} \ee and inserting (\ref{asq}) into (\ref{opia}) one
returns to the results of \cite{mul1}. One would expect that the
effects of confinement suppress amplitude \cite{mul1} in two
different ways: first, because of the mass gap (and actually quite
large mass even of the lightest glueball), and, second, due to the
glueball wave function fast decrease at large relative distances
(see (\ref{f1ree})). These properties solve the artefact of color
Van der Waals forces, appearing in purely perturbative
dipole-dipole interaction.

\section{Conclusions}

In the present paper we discussed interactions of Wilson loops in
confining theory, having in mind gluodynamics as a concrete
example. The effects of confinement were taken into account in the
formalism of perturbation theory in confining background. We have
described the background by gauge-invariant Gaussian correlator
with small correlation length, which is supported by lattice and
analytic calculations. Two main physically different mechanisms of
interaction were analyzed. The first one, which we call
nonperturbative, refers to the process of confining string
rearrangement, which can be energetically preferable for
particular geometries of the contours. In this way a common
surface of two contours $C_1$, $C_2$ is created and in case of
opposite orientation this surface is a ring between $C_1$ and
$C_2$ (with a hole inside the smaller loop). This mechanism has a
direct classical analog in soap films, while for parallel
 orientation nonabelian properties of Wilson loops lead to a
 nonclassical configuration with the same ring but the hole filled
 by the film.
 The second mechanism arises due to two-gluon exchange between loops
 and the corresponding amplitude is $O(g^4)$. In the confining
 background and at large $N_c$ this simple picture of two contours
 connected by two gluon lines is transformed into a new geometry of
 two new composite loops, as shown in Figs.4,5. As a result one has two
 types of surface configurations -- for small and for large separations
  between minimal surfaces $S_1$ and $S_2$, shown respectively in
  Fig.4
  and Fig.5. It is argued in the paper that the configuration
  generic for the
  scattering corresponds to the Fig.5, and reduces to the (reggeized)
  glueball exchanges between loops while for the case of decay and
  sea quark loop effects both nonperturbative and perturbative
  mechanisms are important with small separation between $S_1$
  and $S_2$. Results obtained in the paper provide a basis for a systematic
development both in direction of nonperturbative approach to the
hadron scattering and in the direction of the theory of strong
hadron decays.

\section{Acknowledgments}
The authors acknowledge the support from the grants
RFFI-00-02-17836, RFFI-00-15-96786 and from the grant INTAS
00-110. V.Sh. grateful to the foundation "Fundamenteel Onderzoek
der Materie" (FOM), which is financially
 supported by the Dutch National Science Foundation (NWO).
V.Sh. also acknowledges the support from the grant
RFFI-01-02-06284. The work of another author (Yu.S.) was supported
by DOE contract DE-AC05-84ER40150 under which SURA operates the
Thomas Jefferson National Accelerator Facility, and by the grant
INTAS 00-366.


\vspace{3cm}
\begin{center}
{\Large \bf } \vspace{0.5cm}
\end{center}

\begin{figure}[ht!]
\begin{center}
     \epsfig{file=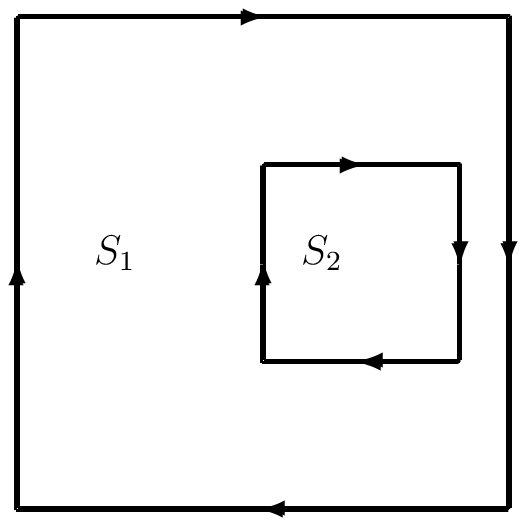, width=50mm}
\end{center}
  \caption[]
  {
   \label{figa1}
Planar geometry of the Wilson loops for $S_2^{min} \subset
S_1^{min}$ (the case of coinciding orientations).
   }
\end{figure}

\begin{figure}[ht!]
\begin{center}
     \epsfig{file=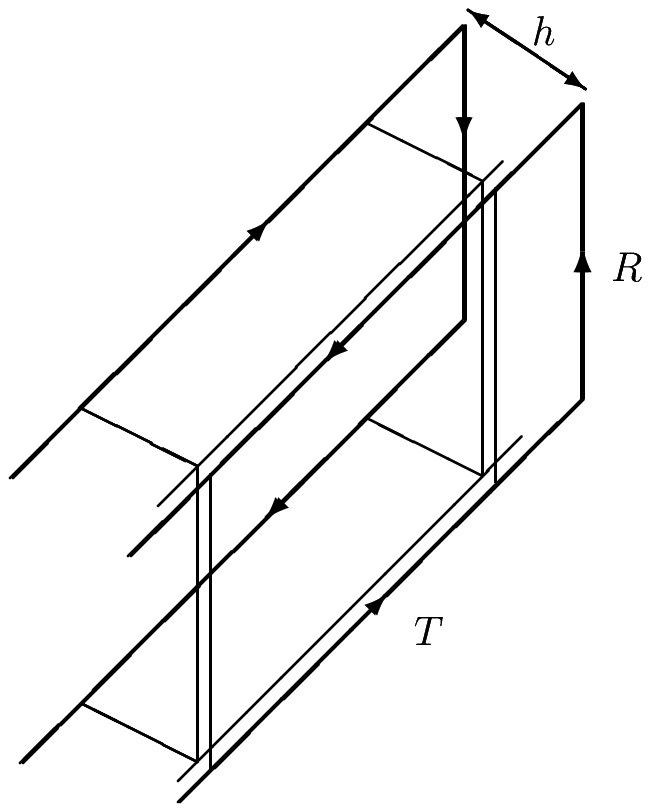, width=60mm}
\end{center}
  \caption[]
  {
   \label{figa2}
Nonperturbative interaction of rectangular Wilson contours. The
minimal string profile for the case of opposite orientations is
given by "enveloping" geometry. Leading large area contribution
corresponds to annihilation of fluxes along $R$.
   }
\end{figure}

\begin{figure}[ht!]
\begin{center}
     \epsfig{file=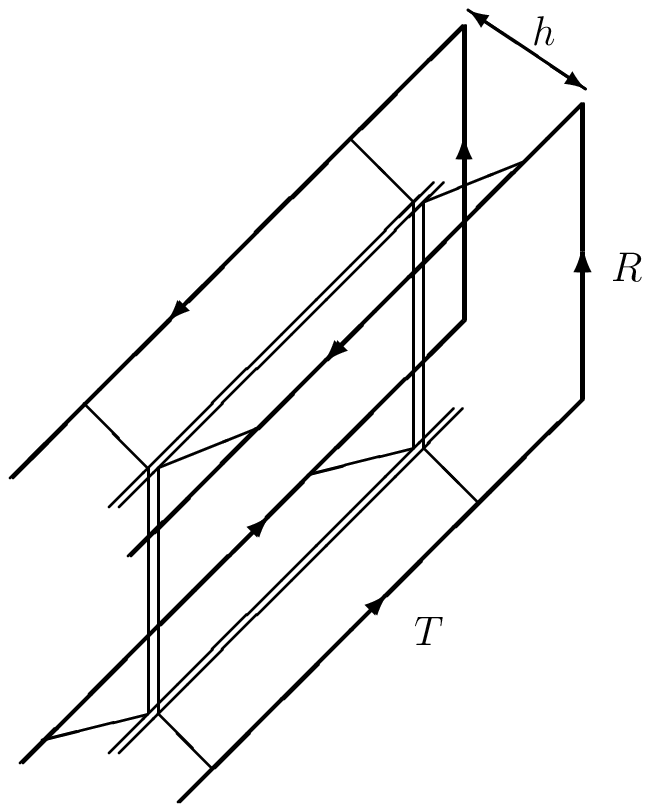, width=60mm}
\end{center}
  \caption[]
  {
   \label{figa3}
 Nonperturbative interaction of rectangular Wilson
contours. The "time-slices" of the minimal string profile for the
case of parallel orientations are depicted. It is assumed that
$N_c = 3$. Leading large area contribution corresponds to a single
fundamental flux in $R$-direction.
   }
\end{figure}

\begin{figure}[ht!]
\begin{center}
     \epsfig{file=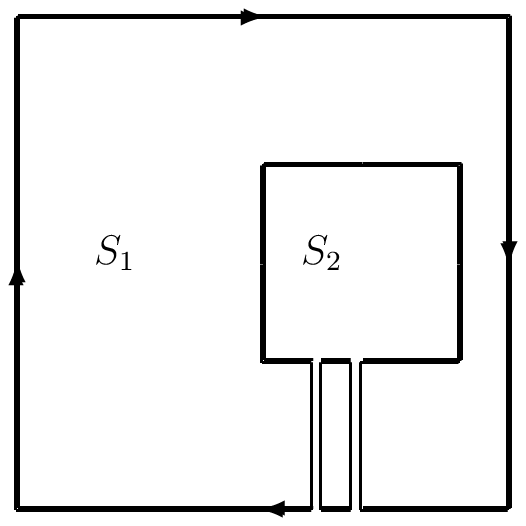, width=50mm}
\end{center}
  \caption[]
  {
   \label{figa4}
The same as Figure 5, but with the two-dimensional geometry of
Figure 1.
   }
\end{figure}
\begin{figure}[ht!]
\begin{center}
     \epsfig{file=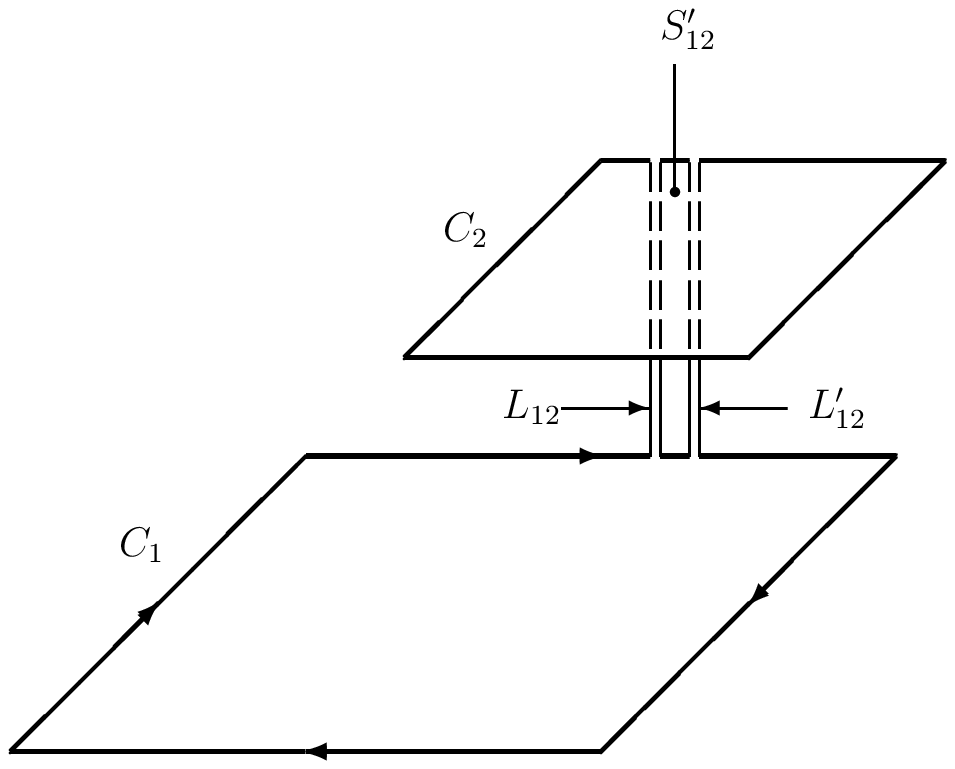, width=75mm}
\end{center}
  \caption[]
  {
   \label{figa5}
Dominant two-gluon glueball term responsible for perturbative
interaction in nonperturbative background at large distances.
Gluon propagator lines are replaced by double fundamental lines in
large $N_c$ limit.
   }
\end{figure}

\end{document}